\documentclass{article}

\usepackage{arxiv}

\usepackage[utf8]{inputenc} 
\usepackage[T1]{fontenc}    
\usepackage{hyperref}       
\usepackage{url}            
\usepackage{booktabs}       
\usepackage{amsfonts}       
\usepackage{nicefrac}       
\usepackage{microtype}      
\usepackage{lipsum}
\usepackage{graphicx}
\usepackage{caption}
\usepackage{enumitem}

\hypersetup{
	colorlinks=true,
	citecolor=red,
	linkcolor=black
}

\title{Predicting next shopping stage using Google Analytics data for E-commerce applications}

\author{
  Mihai Cristian Pîrvu \\
  MorphL, UPB\\
   \And
 Alexandra Anghel \\
  MorphL \\
}

\begin{document}
\setlist{nosep,after=\vspace{\baselineskip}}
\maketitle

\begin{abstract}
E-commerce web applications are almost ubiquitous in our day to day life, however as useful as they are, most of them have little to no adaptation to user needs, which in turn can cause both lower conversion rates as well as unsatisfied customers. We propose a machine learning system which learns the user behaviour from multiple previous sessions and predicts useful metrics for the current session. In turn, these metrics can be used by the applications to customize and better target the customer, which can mean anything from offering better offers of specific products, targeted notifications or placing smart ads. The data used for the learning algorithm is extracted from Google Analytics Enhanced E-commerce\footnote{\url{https://developers.google.com/analytics/devguides/collection/analyticsjs/enhanced-ecommerce}}, which is enabled by most e-commerce websites and thus the system can be used by any such merchant. In order to learn the user patterns, only its behaviour features were used, which don't include names, gender or any other personal information that could identify the user. The learning model that was used is a double recurrent neural network which learns both intra-session and inter-session features. The model predicts for each session a probability score for each of the defined target classes.
\end{abstract}

\section{Introduction}

The internet is crowded with e-commerce businesses, with numbers ranging from 12 to 24 million websites in 2019\cite{top-ecommerce-companies-of-2019}. E-commerce websites are commonly structured around items lists, items details, a shopping cart and a checkout process, with an optional step for paying online. Most of the time, the shopping experience doesn't take into account the user's needs or history. Personalization is employed at a basic level, by profiling users in general categories based on gender, age, location, mobile / desktop device and so on. 

On the other hand, customers receive the same type of notifications and offers, with little or no attention to their specific requirements. General discount rates or rule-based calls to action are being sent, such as email notifications for finalizing a purchase when an item was already added to the cart. 

Using this general approach means missing out on a huge opportunity to increase conversion rates by engaging users with a personalized experience. Clearly, if an e-commerce application wants to maximize its profits, it should adapt to the user, rather than just offering a one-size-fits-all approach.

In this article, we present a learning algorithm which predicts the outcome of a user's browsing session, using the information gathered from his previous sessions as well as the data that the user provides during the current one. The data that the user generates is gathered from the Google Analytics Reporting API v4\footnote{\url{https://developers.google.com/analytics/devguides/reporting/core/v4}}. The same model can be applied to Google Analytics 360 / BigQuery.

Google Analytics is very useful for generating activity reports or segmenting users. In addition, the Enhanced E-commerce section offers an overview of the sales funnel. However, this funnel doesn't offer any insights into why some users convert and some do not.

Our proposed method enables merchants to enhance the user experience by getting to know beforehand the behaviour of the user during a particular session. The behaviour is defined as a probability vector for the most common actions that a user can take: visit a regular page, visit an item details page, add an item to the online cart, visiting the checkout page and actually making a successful transaction. These actions correspond to the purchase funnel\footnote{\url{https://support.google.com/analytics/answer/6014872?hl=en\#sba}}, available as a report in the Google Analytics dashboard. For each of these classes, based on the user's history and user's features in the current session, a probability vector is outputted. This vector doesn't include the value of the predicted transaction, just the probability that the user will make a purchase. The same applies for all classes. It becomes straightforward that the retailers can use this information to target the user with better ads or notifications, or even offer incentives, such as gifts or discounts, while allowing the shop to also make a profit. This becomes a win-win situation, using only the data at hand.

In Section \ref{sec:data-processing} we will present how the data is exported, what features are kept and how they are processed in order to enable learning. Then, in Section \ref{sec:model-architecture}, we will present the learning algorithm, the architecture of the model and other information related to training. Lastly, in Section \ref{sec:results}, we will provide quantitative and qualitative results as well as present an experiment on real world data in order to compare the results of using the trained model against pure statistical solutions.

\section{Related Work}

In the domain of website personalization via learning, two classical approaches are usually taken: learning from the data alone using various regression or classification models, such as linear regression, SVMs or neural networks. These methods can also be extended to include larger time lines as context, resulting in recurrent models. The second direction is using reinforcement learning, where the goal is optimizing some non-differentiable criteria, such as maximizing number of clicks on an ad, maximizing profit or minimizing churning.

For Recommender Systems, classical approaches use either collaborative filtering or content-based filtering using supervised learning. Matrix factorization \cite{koren2009matrix} is a prime example of collaborative filtering, however the main problem is that it can suffer from cold-start issue with new users. This is partially solved using recurrent neural networks on the data alone, such as in \cite{tan2016improved}.

Another classical problem is the task of Intent Prediction, which aims at finding the intention of a user during the current session. Recent methods include predicting if the user will buy during the current session using recurrent neural networks \cite{sheil2019discovering}. Other approaches involve splitting the intent of a user in multiple disjoint classes, such as informational, transactional, considerational or navigational, corresponding to the marketing funnel.In \cite{pirvu2018predicting}, we have predicted these classes using only the search query that the user employed to land on the page. We used a semi supervised approach, by first creating an auto-labeling process and annotating a large amount of queries from a big corpus and train a partial model. Then, we fine tuned the model using a small subset of queries that were manually labeled, which proved better than using any of the steps individually.

One of the most pioneering works in the field of reinforcement learning for website personalization is using contextual bandits \cite{li2010contextual, bietti2018contextual}. This algorithm tries to optimize a given criteria, say number of clicks on an ad, in a one-action system using the historical data for each action. The classical non-contextual bandits only use the statistics about previous clicks and then, after computing scores for each action and using a choice strategy (which is a classical example of exploration vs. exploitation), an ad will be shown. The contextual bandits extend this problem to adding user features in the mix, such as previous page, user history, how the user arrived to the current page and so on. This kind of features is what our learning algorithm tries to leverage as well, but we are limited to the features offered by Google Analytics.

Using Google Analytics data for Machine Learning purposes is a relatively new and rarely used method, perhaps because the analytics platform's original focus is on aggregated data. The User Explorer\footnote{\url{https://support.google.com/analytics/answer/6339208?hl=en}} report, which provides insights at the user level, is relatively new. In \cite{gunter2016forecasting}, they predict the amount of people visiting a city based on website traffic of various touristic websites. In \cite{durden2016identifying}, they try to identify the demography of the users on a website, studying the evolution of location as well as device used for some time frames. In \cite{vecchione2016tracking}, they analyzed the website traffic to sort the most visited pages for some online library. Then, based on the traffic analysis, they put more emphasis on the most visited ones, promoting them to the main page more often which resulted in a decrease of the bounce rate by a large margin. In \cite{plaza2011google}, they analyze the performance of various e-commerce sites using computed statistics of the features offered by Google Analytics.

We can see that most of these articles focus on human made assessments and simple statistics, while our method uses the data to train a recurrent neural network which in turn is used to provide a competent insight that can be used by the website programmatically in order to improve its performance.

\section{Data Processing}
\label{sec:data-processing}

The learning algorithm leverages on data that is readily available to any e-commerce website that has enabled Google Analytics Enhanced E-commerce. However, in order to implement a proof of concept for the task of predicting the next shopping stage, we took a small sample of a predefined time frame from an online e-commerce retailer. All the experiments and statistics that are computed are only valid for that particular website. Therefore, in order to apply the same learning algorithm on a new website, the statistics must be recomputed and the model must be retrained and fine-tuned for it.

The Google Analytics Reporting API v4 enables administrators to export its logged data in JSON format, which can be processed, turned into a tabular form with numerical features and used for training machine learning methods. The data is divided into three main components: user data, session data and hits data. User data represents the information about each user, meaning an unique Client ID, device (mobile or desktop), browser and user type (New vs. Returning Visitor). It should be noted that the Client ID refers to a browser, not to a user account, thus it doesn't contain any personal data. It is possible to associate the Client ID with a user account (across devices) by providing an authentication feature in the E-commerce application, however, in this particular use case, all client ids refer to browsers and have no correlation with real life names or any information that could be used to identify the user. Session data contains the information about each session, such as duration, number of transactions, number of searches, etc. Finally, hits data represent the intra-session information, such as the time on each page during the current session, the time when the page was accessed or the fact that the user looked at details of a product. 

The basic user and session data is generally available for all websites that include the Google Analytics script, without additional setup. However, E-commerce shopping stages, such as visualizing items lists, items details, adding a product to the cart and the checkout process require additional code that must be added by the webmaster. These actions are then sent to Google Analytics via the DataLayer. The setup process is described at length in the "Enhanced E-commerce (UA) Developer Guide"\footnote{\url{https://developers.google.com/tag-manager/enhanced-ecommerce}}. In addition, in order to allow exporting the user-level, session-level and hits-level data via the Google Analytics API, custom dimensions must be added, uniquely identifying each user / browser, each session and each hit / timestamp.

\subsection{Features Description}

Next, we'll talk about each feature in particular, giving a small description about them and try to understand how they influence the decision of the user at the end of each session. For a complete list of features included in the Google Analytics API, one can visit the official page \footnote{\url{https://developers.google.com/analytics/devguides/reporting/core/dimsmets}}.

\textbf{User features.}\hspace{0.3cm} The user features are invariant to each user. If the same person uses two browsers or devices, then he or she will be counted as two different users by this system.

\begin{tabular}{|c | l|} 
 \hline
 Feature & Description \\ [0.5ex]
 \hline
ClientID & Unique Client ID in Google Analytics, used to track the user across multiple sessions \\
& using browser cookie. It must be set as a custom dimension in order to export user data from the API. \\
 \hline
 User Type & Whether this user is a new visitor or a returning visitor \\
\hline
Device Category & The category of the device: mobile, tablet or desktop \\
\hline
Browser Name & The raw name of the browser \\
\hline
Browser Revenue & Browser feature computed from the amount of revenue of each \\
per Transaction & browser divided by the total amount of transactions done using that browser \\
\hline
Device Name & The raw name of the mobile device\\
\hline
Device Revenue & Device feature computed from the amount of revenue of each \\
per Transaction & device divided by the total amount of transactions done using that device \\
\hline
\end{tabular}

The ClientID is used for joining together the tables. Browser Name and Device Name cannot be used as features, because this is a categorical feature that can grow indefinitely. However, by including it, we can compute a statistical number regarding how much revenue a transaction brings using each particular browser and device. This is computed by summing all the transaction revenues using each mobile or browser and then dividing by the count of the transactions for that item. A statistical analysis for these features can be seen in Figures \ref{fig:browser_features} \& \ref{fig:device_features}.

\noindent%
\begin{minipage}{\linewidth}
\centering
\makebox[\linewidth]{
\includegraphics[scale=0.24,keepaspectratio]{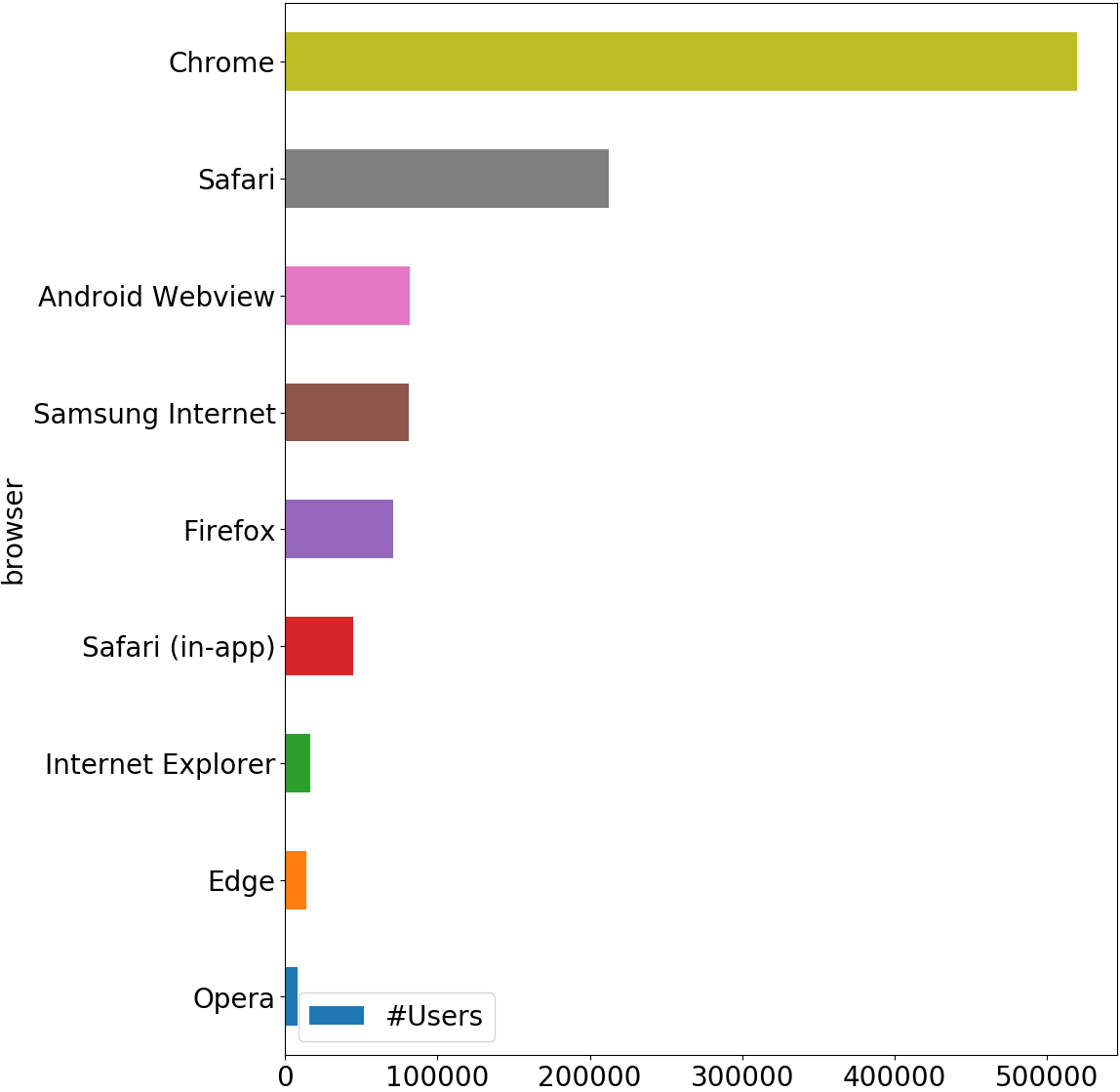}
\includegraphics[scale=0.24,keepaspectratio]{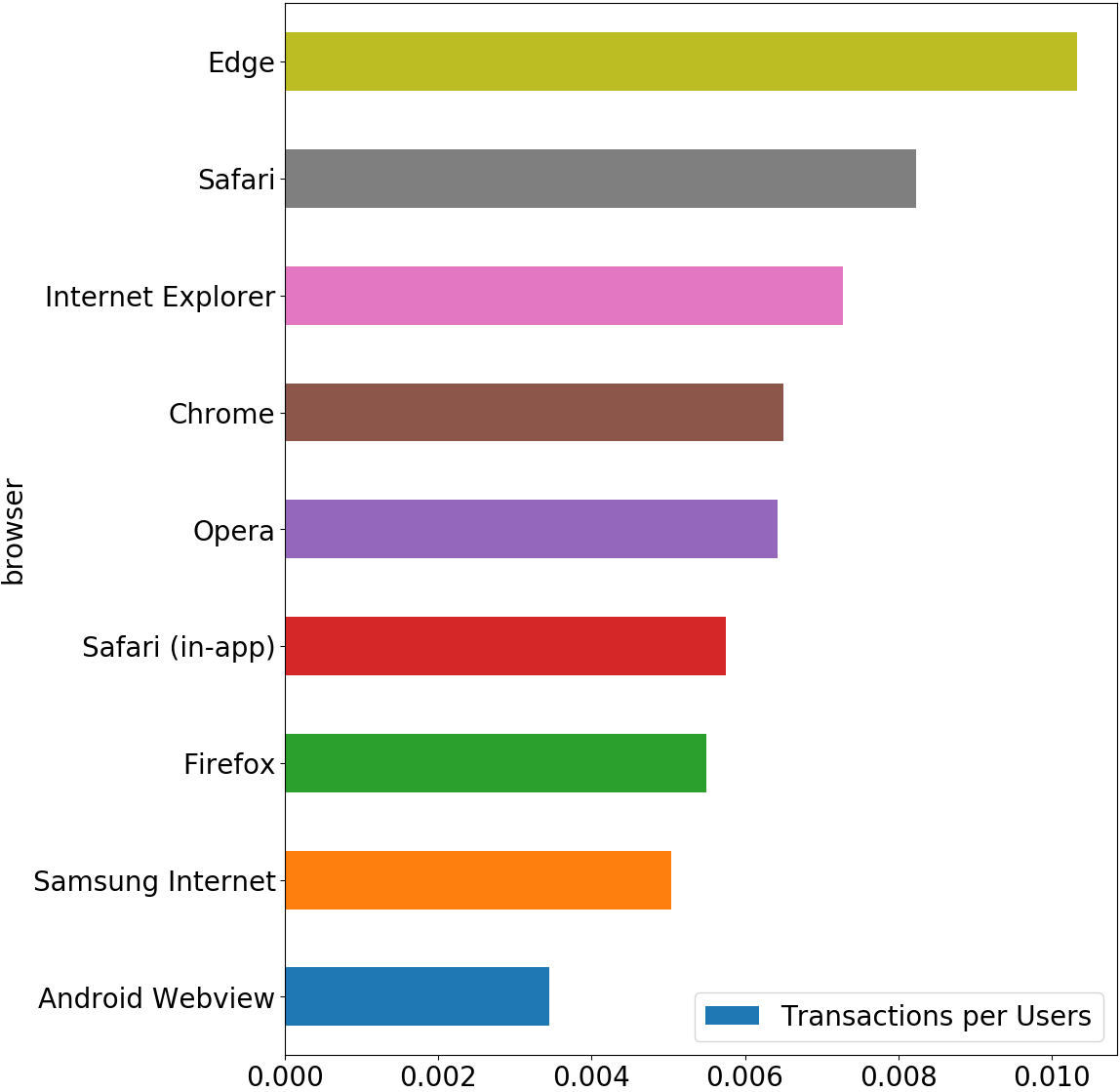}
\includegraphics[scale=0.24,keepaspectratio]{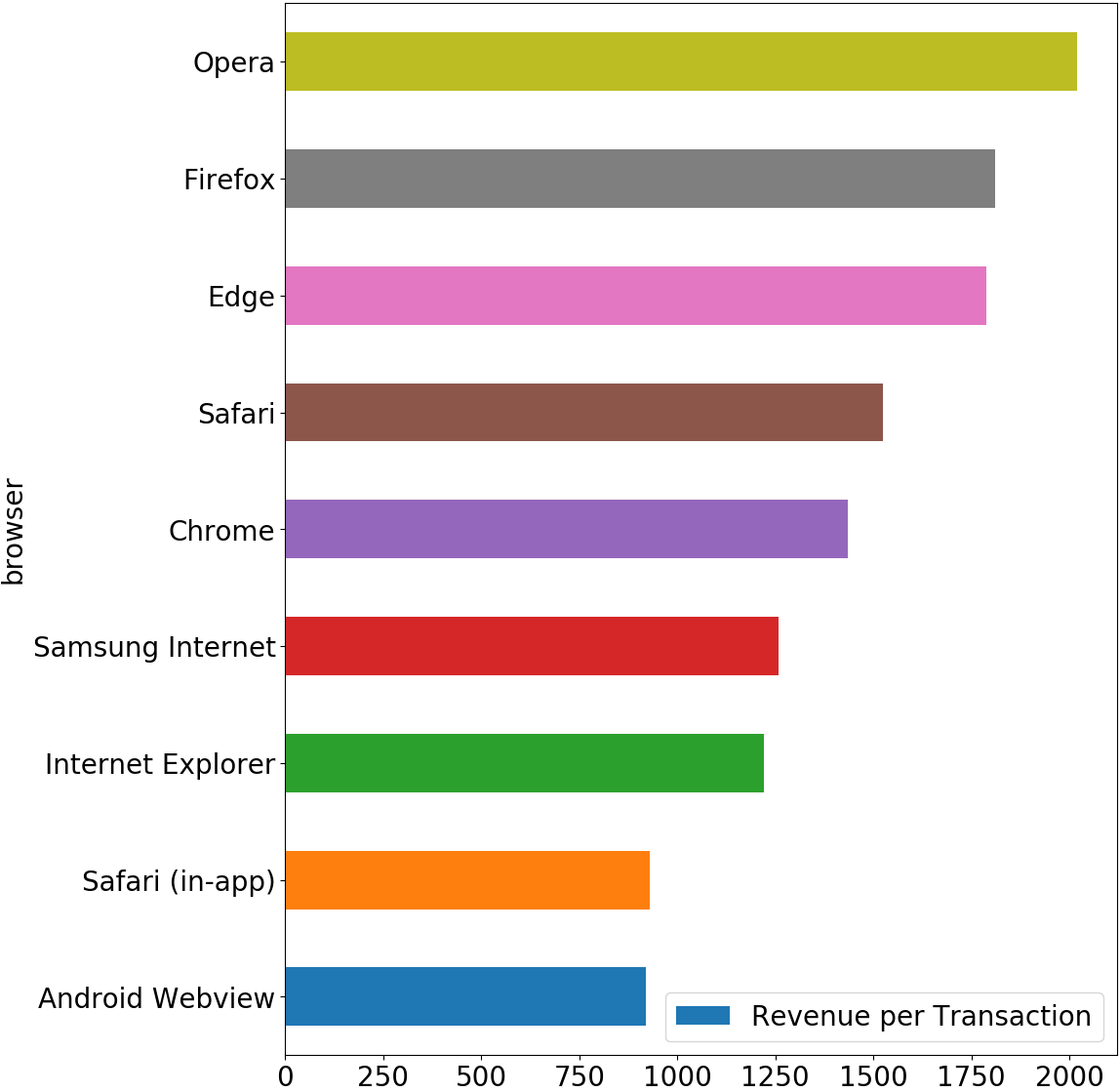}
}
\captionof{figure}{Users histograms based on browsers}\label{fig:browser_features}
\end{minipage}

\noindent%
\begin{minipage}{\linewidth}
\centering
\makebox[\linewidth]{
\includegraphics[scale=0.26,keepaspectratio]{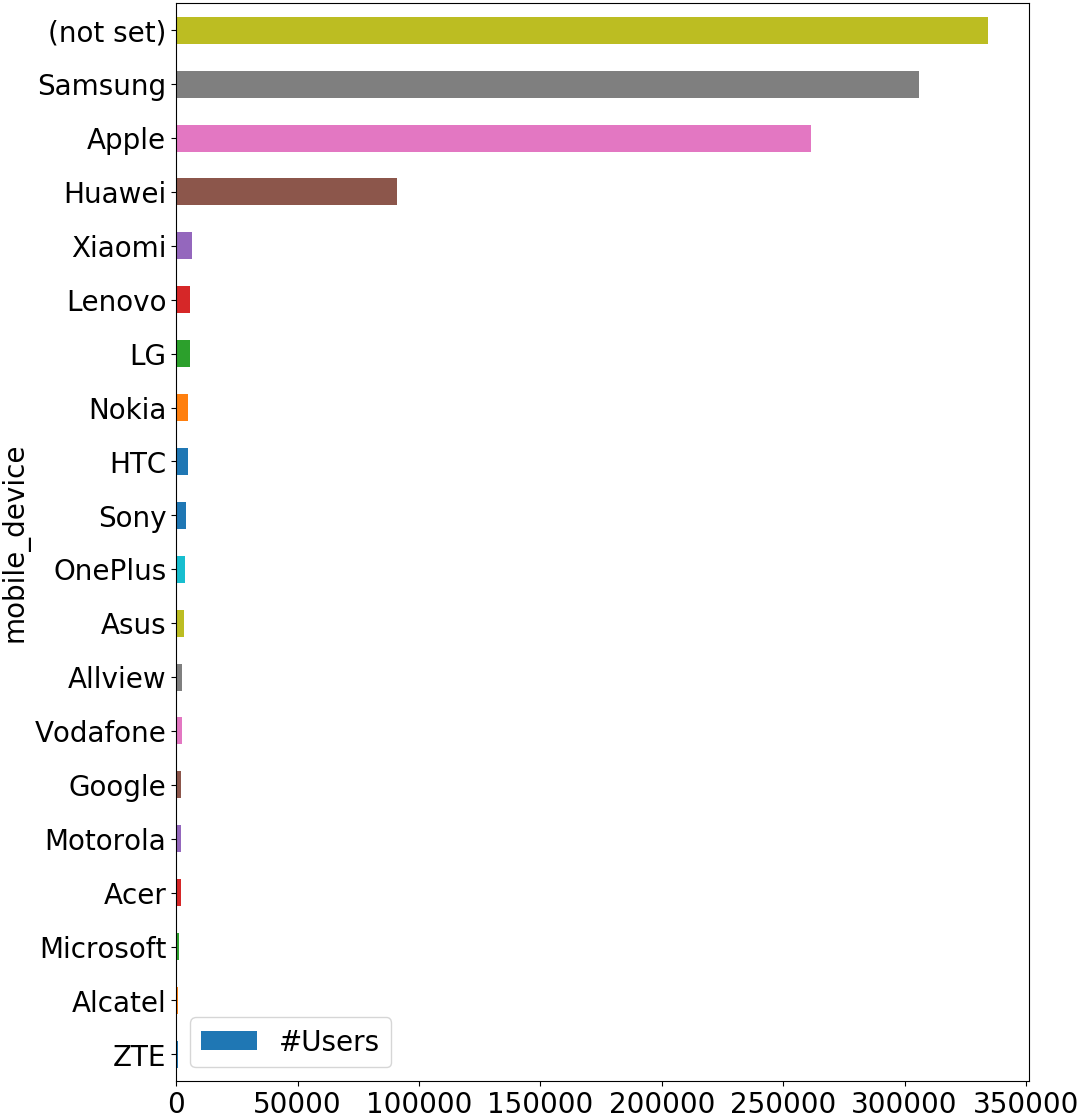}
\includegraphics[scale=0.26,keepaspectratio]{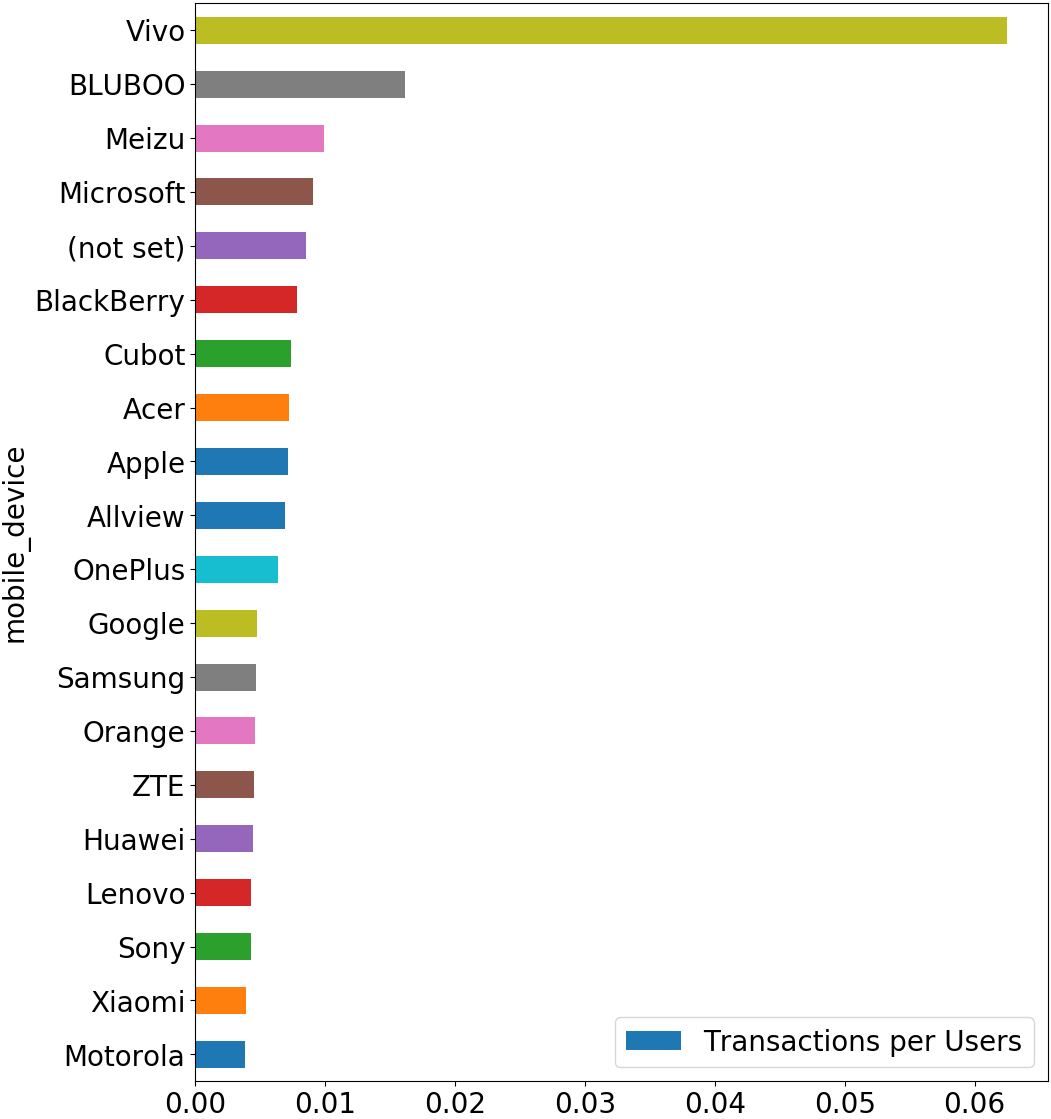}
\includegraphics[scale=0.26,keepaspectratio]{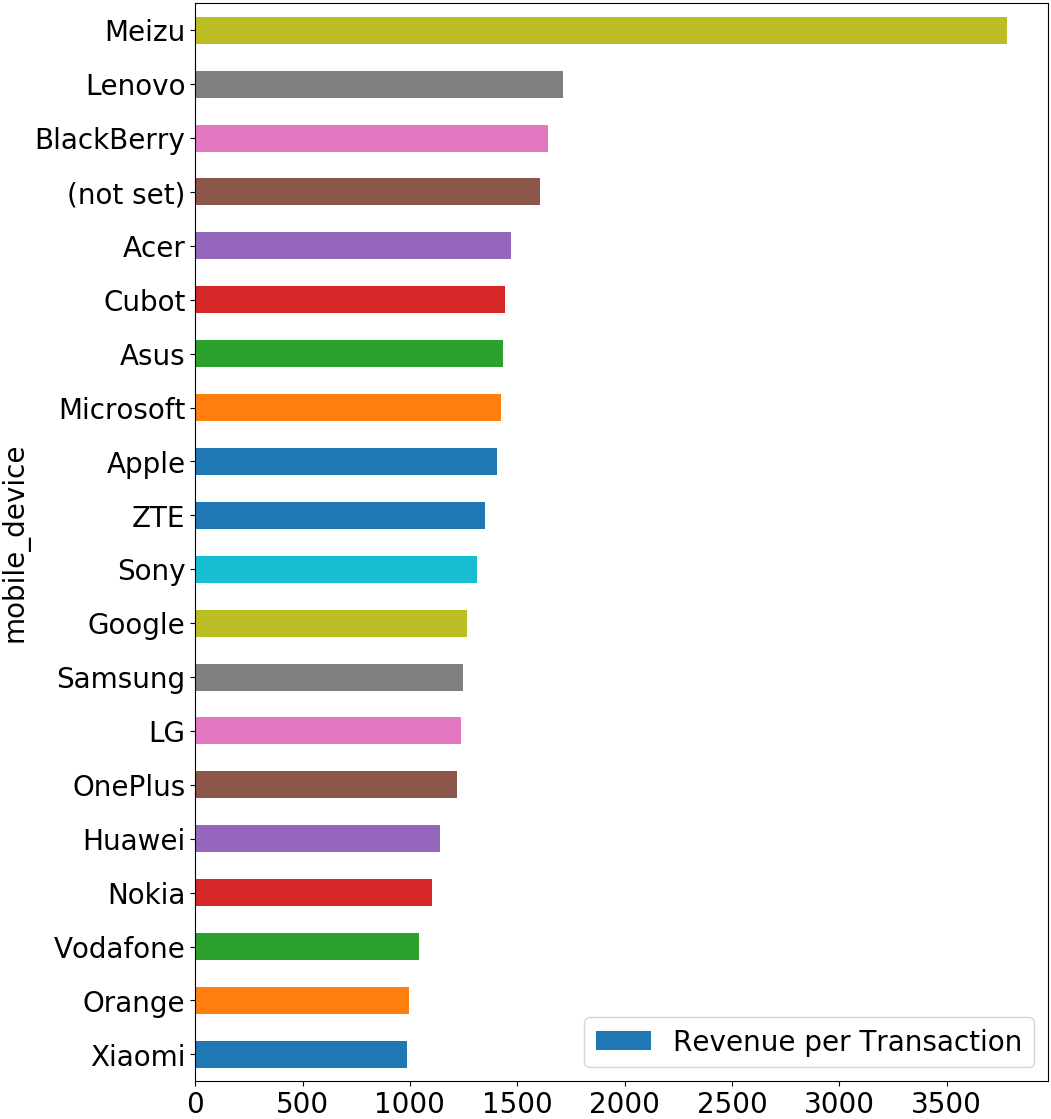}
}
\captionof{figure}{Users histograms based on devices}\label{fig:device_features}
\end{minipage}

We can see that while the number of users using a particular device or browser is dominated by a small amount of items, the actual revenue per transaction differs vastly. The final feature we include from this insights is the last histogram, where the raw number if used, thus prioritizing high selling devices over low selling ones.

\textbf{Session features.}\hspace{0.3cm} The session features are those which remain invariant during a session, as opposed to hits features, which change for each page view.

\begin{tabular}{|c | l|} 
 \hline
 Feature & Description \\ [0.5ex]
 \hline
ClientID & Unique Client ID in Google Analytics, used to track the user across multiple sessions \\
& using browser cookie \\
\hline
SessionID & Unique Session ID in Google Analytics. Set as a custom dimension,\\
& composed of a random string and a timestamp. \\
\hline
Session Duration & The duration of the current session \\
\hline
Unique Pageviews & Amount of unique pages that were visited during the current session \\
\hline
Transactions & Number of transactions that took place during this session \\
\hline
Revenue & The amount of money spent\\
\hline
Unique Purchases & The amount of unique items that were purchased \\
\hline
Days Since Last Session & Integer representing the number of days between the previous session and current one \\
\hline
Site Search Status & Boolean representing whether the internal website search function was used \\
\hline
Results Pageviews & Number of times the resulting page of an internal search was accessed \\
\hline
Total Unique Searches & Total number of unique internal searches. If the same keyword is searched multiple \\
& times, it will be counted only once \\
\hline
Search Depth & The total number of subsequent page views made after a use of the site's \\
& internal search feature. \\
\hline
Search Refinements & The total number of times a refinement (transition) occurs between \\
& internal keywords search within a session. \\
 \hline
 Shopping Stage & The shopping stage at the end of each session. This is the target feature for \\
  & this use case.\\
\hline
\end{tabular}

The first two features are only used to join together the hits and user features. However, all the other columns are transformed into numerical values and used as-is in the learning process. It may not be obvious that all columns are useful for predicting the next shopping stage, however some of them, such as revenue or number of transactions definitely influence. It should be noted that for predicting the current shopping stage, the features of the previous session are used directly, because otherwise the transactions or revenue columns would directly tell us if a user made a payment or just visited the website. This also enables the system to be used as a real-time application, where the features of the previous sessions are used in addition to the hits of the current session in order to predict the shopping stage of the current session. More about this will be detailed in Section \ref{sec:model-architecture}.

\textbf{Hits features.}\hspace{0.3cm} The hits features are those which are updated for each page view of each session.

\begin{tabular}{|c | l|} 
 \hline
 Feature & Description \\ [0.5ex]
 \hline
ClientID & Unique Client ID in Google Analytics, used to track the user across multiple sessions \\
& using browser cookie \\
\hline
SessionID & Unique Session ID in Google Analytics \\
 \hline
 Date Hour and Minute & The hour when the hit was made \\
 \hline
 Time on Page & The amount of time the hit took \\
 \hline
 Product Detail Views & Whether the user looked at the details page of a product \\
 \hline
\end{tabular}

We can see that there are only a few usable features, which is a downside of using Google Analytics in its current form. However, if more hits were to be enabled, this would only increase the quality of the data and, in turn, the quality of the results. In Figure \ref{fig:correlation_matrix}, we can see a correlation matrix for all the features using Pearson's correlation coefficient.

\noindent%
\begin{minipage}{\linewidth}
\makebox[\linewidth]{
  \includegraphics[scale=0.25,keepaspectratio]{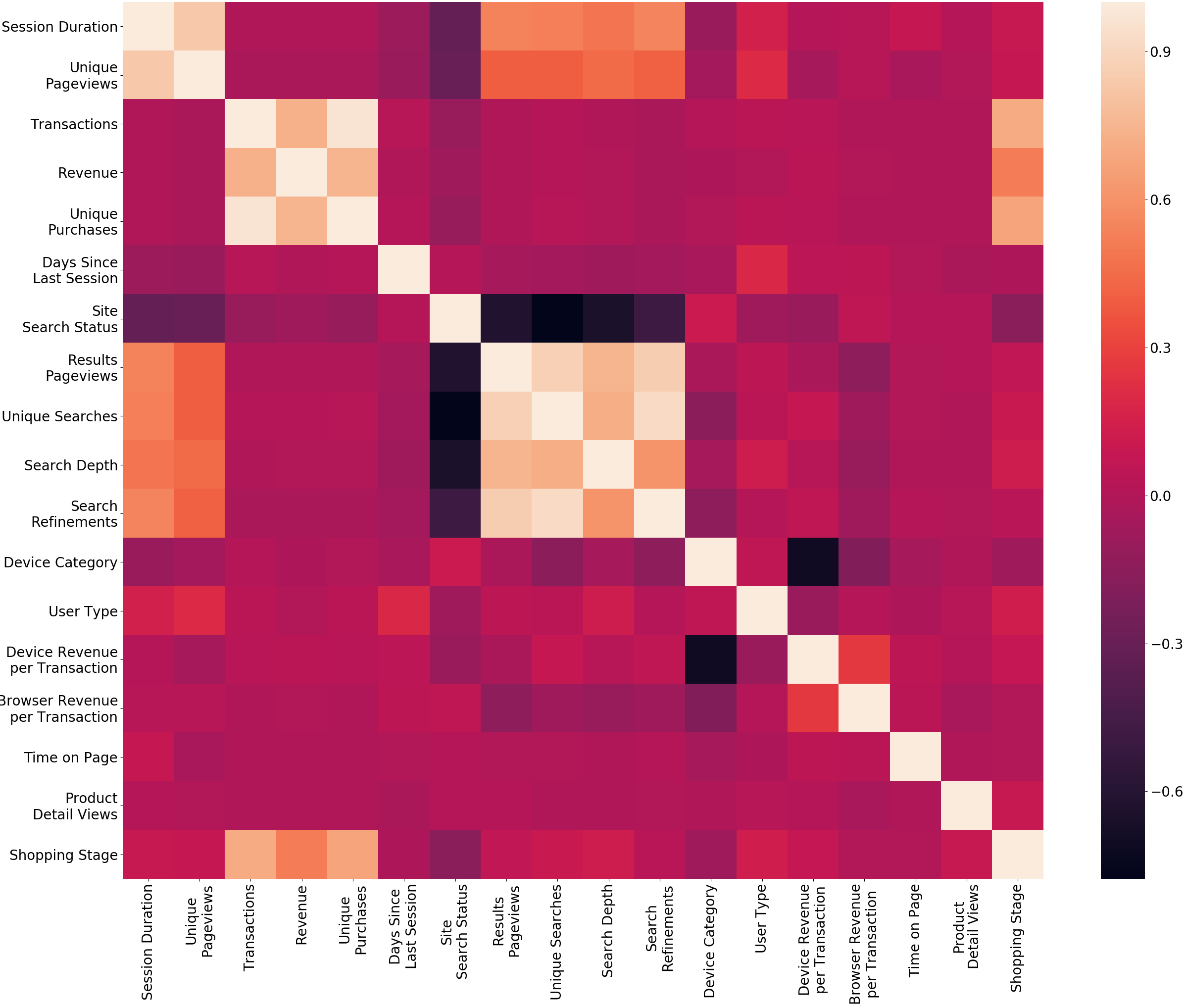}
}
\captionof{figure}{Features correlation matrix}\label{fig:correlation_matrix}
\end{minipage}

The last column is the target column and we can see that most if not all features have some correlation to it. The strongest ones are number of transactions, revenue and unique purchases in the session. This is the reason why we're concatenating the features of the previous session to the current session instead of using the current ones. We'd have no idea how many transactions a user has made at the start of a session in a real life situation. Also, those features hide the shopping stage. It is also important to observe that all features provide some sort of correlation, meaning that, while this problem is starved of features, all of them are somewhat relevant. One final thought is that adding more custom features would increase the quality of the result.

\subsection{Modeling the Ground Truth}

Our initial intention was to simply predict the shopping stage as a simple classification problem, predicting one of the 6 possible classes. One issue that we had to solve was that Google Analytics doesn't offer us the real time shopping stage, which means that for each hit (page view), we get the status of the user. Instead, it offers us all the shopping stages the user went through at the end of the session, without knowing which hit caused which stage. Therefore, after removing outliers and aggregating unique paths, we are left with the following classes: 

\begin{itemize}
\item {All Visits}
\item {All Visits -> Product View}
\item {All Visits -> Product View -> Add to Cart}
\item {All Visits -> Product View -> Add to Cart -> Checkout}
\item {All Visits -> Product View -> Checkout}
\item {Transaction}
\end{itemize}

These are the 6 disjoint and standard paths a user usually takes when visiting an E-commerce website. The first one corresponds to simply visiting a misc page, such as contact page information about the retailer. The second one is the opposite, where the user visits a page that contains a product or the list of all products. The next ones correspond to the user adding an item to the shopping cart and checking out a shopping cart. Finally, the last one represents the most important one for this use case, where a user has successfully bought a product.

However, we quickly realized that, because of data imbalance, the classifier would always predict the most dominant classes (first two), so we changed the problem formulation from a classification problem to a regression problem. Now, the model has to predict a probability for each of the 6 classes. However, the main problem now was deciding how to model this probability, as we'd like it to be as close to the actual probability of a user making a Transaction, or making an All Visits hit and so on. Our solution was to compute it using two attribution modelings \footnote{\url{https://support.google.com/analytics/answer/1662518?hl=en}}, namely linear and time decaying.

\subsubsection{Linear attribution modeling}
\label{subsubsec:linear-attribution-model}
This attribution model simply gives equal contribution for a transaction to all sessions, regardless of the time that passed between the first and last sessions.

Let's consider a user has 5 sessions and two of these are transactions. We can view this as a binary vector: $ v = [0, 1, 0, 0, 1] $. Using a linear attribution model, this user has a transaction probability of $2 / 5 = 0.4$. However, since we are dealing with a recurrent neural network, we'd like to learn from this user his whole behaviour, starting from his first session up until his fifth. Thus, his transaction probability up until his n-th session is the partial sum of his transactions starting from the first session all the way to the nth one. Formally, $ t(n) = \frac{\sum_{i=1}^{n}{v(1:n)}}{n} $, for all $n \leq N$, where N is the total number of sessions for this user and $:$ is the partial vector operator. The resulting vector t for this particular user is $ t = [\frac{0}{1}, \frac{1}{2}, \frac{1}{3}, \frac{1}{4}, \frac{2}{5}] = [0, 0.5, 0.33, 0.25, 0.4]$.

This is the ground truth vector that the recurrent model must learn at each time step. The same process is repeated for all 6 classes, not just transactions. We can see that all the sessions indeed contribute equally, giving a large value to the transaction that was made in the 2nd session all the way up until the 5th session. The next attribution model changes this, by giving more weight to the sessions that are closer to the last session.

\subsubsection{Time decaying attribution modeling}

The second attribution modeling tries to give more value to the latest sessions, by applying a half-life weight based on how far away the session is. Basically, given the same example as before, if we have the transactions vector $ v = [0, 1, 0, 0, 1] $, then the weight vector would be $ w = [\frac{1}{2^4}, \frac{1}{2^3}, \frac{1}{2^2}, \frac{1}{2^1}, \frac{1}{2^0}] = [0.06, 0.12, 0.25, 0.5, 1] $. Making a parallel to the linear case, the weight vector there is simply 1 for each position. 

Then, the partial transaction probability value can be formulated as: $ t(n) = \frac{w(1:n) \cdot v(1:n)}{\sum_{i}^{n}{w(1:n)}} $, for all $ n \leq N $ where $ \cdot $ is the dot product between two vectors. Technically, the same formula can be applied to the linear case, however the denominator sums up to n, because the weight vector contains just ones and the numerator simplifies to the sum above.

For this particular case, the transaction probability vector is:
$$
t = [\frac{0 * 1}{1}, \frac{0*\frac{1}{2} + 1 * 1}{\frac{1}{2} + 1}, \frac{0*\frac{1}{4} + 1*\frac{1}{2} + 0*1}{\frac{1}{4}+\frac{1}{2}+1}, \frac{0*\frac{1}{8}+1*\frac{1}{4}+0*\frac{1}{2}+0*1}{\frac{1}{8}+\frac{1}{4}+\frac{1}{2}+1},\frac{0*\frac{1}{16}+1*\frac{1}{8}+0*\frac{1}{4}+0*\frac{1}{2}+1*1}{\frac{1}{16}+\frac{1}{8}+\frac{1}{4}+\frac{1}{2}+1}] = $$
$$ = [0, 0.66, 0.28, 0.13, 0.55]$$.

We can see that this method puts much more value on the more current ones. Had the first transaction moved 1 session closer to the last one, then the transaction probability value of the last session would've become 0.65 according to the same formula.

\section{Model Architecture}
\label{sec:model-architecture}

The model proposed to tackle this problem is a double recurrent neural network, where both recurrent layers are implemented using LSTM cells. The high level architecture can be seen in Figure \ref{fig:architecture}.

\noindent%
\begin{minipage}{\linewidth}
\makebox[\linewidth]{
  \includegraphics[scale=0.4,keepaspectratio]{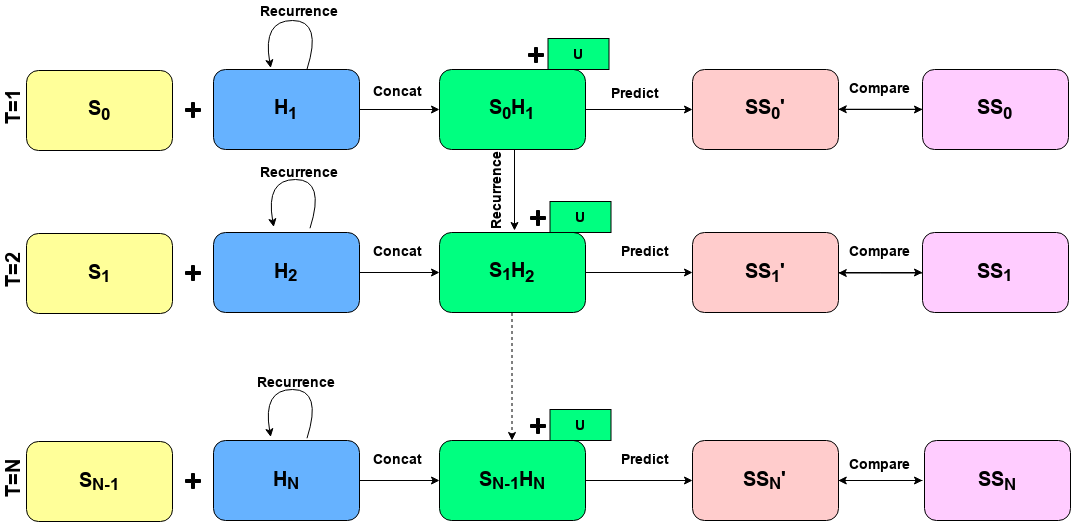}
}
\captionof{figure}{Model Architecture}\label{fig:architecture}
\end{minipage}

The data flow is that, for each user, we divide the entries of the user based on all his sessions. Now, for each session of this user we have a variable number of hits, which is denoted with the blue box in the figure. After passing all the hits features through the first LSTM, we get a hidden state vector, noted as $H_i$, for each session. This feature vector is concatenated with the features of the previous session $S_{i-1}$, with the special case that the first session is concatenated with a zeroed vector, as there is no history beforehand, resulting the feature vector of sessions and hits. Now, all these session + hits vectors are passed through the second LSTM, resulting another feature vector $S_{i-1}H_{i}$ that combines both the features of the current session with the history of the previous sessions, for all sessions, denoted as green in the Figure. For all these feature vectors, we also concatenate the user features and then pass them through two fully connected layers, after which we get 6 probabilities, representing the probability of each possible output, at every session. The activation function for the first FC layer is ReLU, while the second one is the element-wise sigmoid. This is denoted by pink in the Figure. These outputs are compared with the ground truth probability vectors, computed as described earlier depending on the chosen attribution model, using the MSE as the loss function of the model.

The only hyperparameter used for this model is the number of hidden units of the LSTM hidden states, which was empirically chosen as 30, due to the low feature space of the hits. The total number of trainable parameters for this model is just 15.246, which can be considered very lightweight compared to other deep learning models.

\section{Results}
\label{sec:results}

The experiments were run on a private dataset from an E-commerce website, which were exported exactly as described in Section \ref{sec:data-processing}. Both attribution models were used (linear and time decaying), as well as two normalization methods (min max normalization and data standardization).  The results were pretty similar, as we can see in Table\ref{tab:results}. The model was implemented using PyTorch, using only standard layers for the neural network.

\begin{minipage}{\linewidth}
\makebox[\linewidth]{
    \begin{tabular}{|c|c|c|c|c|} 
    \hline
     Attribution model & Normalization & Loss & Accuracy (0.5/2.0) & Accuracy (0.8/1.25) \\ [0.5ex]
    \hline
    Linear & Min-Max & \textbf{0.0056} & \textbf{47.75\%} &\textbf{40.72\%} \\
    \hline
    Linear & Standardization & \textbf{0.0056} & 47.10\% & 39.50\% \\
    \hline
    Time decaying & Min-Max & 0.0065 & 44.73\% & 13.43\% \\
    \hline
    Time decaying & Standardization & 0.0065 & 44.68\% & 13.39\% \\
    \hline
    \end{tabular}
}
\captionof{table}{Training results (loss and accuracies) on the evaluation set.}
\label{tab:results}
\end{minipage}

The accuracies were computed by renormalizing the output, which, for each user represents the percentage of making each of the 6 probable actions. Take, the linear attribution model, which was detailed in Subsection \ref{subsubsec:linear-attribution-model}. If a user has 3 transactions, in 5 sessions, his linear score will be $ 3 / 5 = 0.6 $. The renormalization scheme is done, such as we can defined interval thresholds based on the real number of transactions and the predicted one. Here, at the 5th session, the user has made 3 transations. Thus, for a $ 0.5 \leq x \leq 2 $ interval, we say that the result is correct if the predicted transactions are more than half (1.5) and less than twice (3). This process is only evaluated for the last session, based on the output probability, and the number of sessions of the user. A similar logic is applied to time decaying as well. 

We can observe that the linear model learns much better than the time decaying one. Perhaps, because applying an exponential law to the distance between sessions as a label target is too complicated given the small amount of features that we could gather using Google Analytics. In what follows, we take these models and try to apply it to a real life situation, where we'd like to target users, with an ad, or a promotion or anything else targeting might mean.

\subsection{Targeting Experiment}

Given a subset of the validation set, with a segment of 5347 users, we want to target as many of these users in the last session (which can be thought of the active session). We know that only 32 users performed a transaction in the last session, however we have the history of transactions of all of them in all of their sessions. These 32 transactions accounted for a revenue of 29579 (in some currency). We will analyze four targeting methods, two based on the two machine learning models, one using a random scheme and one using a statistical method similarly to the linear attribution model.

We want to see, for each method, what are the percentiles of true positives (correctly targeted users) and false positives (incorrectly targeted users). In real life situations, each targeted user costs an amount of revenue (the cost of an ad, for example) and the breaking point would represent the maximum amount of revenue that can be spent while also making profit.

Our experiment can be described as follows: Given each user and its history (sessions, hits, transactions, etc.), we compute a transaction probability $0 \leq p_u \leq 1$. Then, for this user we apply a binomial sampling and if the result is positive we target him. Targeting has a cost, but that cost can be alleviated if the user makes a transaction after he was targeted. If not, then the user was targeted for nothing, and we lose money from the total revenue. Thus, given 4 methods: random, statistical, time decaying and linear, we'd like to know how many users (in average) we correctly target (true positives), how many users we wrongly target (false positives).

The random method gives a random probability to all users without any bias. The statistical method is a replica of the linear attribution model, where we use all the sessions besides the last one for each user and apply a random uniform noise of 0.1. The last two methods are the values reported by the trained models. In Table \ref{tab:experiment-results}, we can see the results of this experiment, for this particular validation set.

\begin{minipage}{\linewidth}
\makebox[\linewidth]{
    \begin{tabular}{|c|c|c|c|c|c|} 
    \hline
     Method & True positives & True positives\% & False positives & False positives\% & Breaking cost point \\ [0.5ex]
     \hline
    Random & \textbf{22.46} ($\pm$ 3.28) & \textbf{70.19\%} ($\pm$ 10.24\%) & 2133 ($\pm$ 39.09) & 40.14\% ($\pm$ 0.74\%) & 10.88 ($\pm$ 1.52) \\
     \hline
    Statistical & 2.04 ($\pm$ 1.72) & 6.38\% ($\pm$ 5.38\%) & 97.94 ($\pm$ 8.91) & 1.84\% ($\pm$ 0.17\%) & 12.05 ($\pm$ 9.73) \\
     \hline
    Time decaying & 1.96 ($\pm$ 1.20) & 6.12\% ($\pm$ 3.75\%) & 72.16 ($\pm$ 7.97) & 1.36\% ($\pm$ 0.15\%) & 22.90 ($\pm$ 10.18) \\
     \hline
    Linear & 5.02 ($\pm$ 2.53) & 15.69\% ($\pm$ 7.92\%) & \textbf{36.72} ($\pm$ 6.75) & \textbf{0.69\%} ($\pm$ 0.13\%) & \textbf{103.72} ($\pm$ 24.13) \\
    \hline
    \end{tabular}
}
\captionof{table}{Experiment results, in term of real values as well as percentiles for true positives, false negatives and revenue per targeting breaking cost point.}
\label{tab:experiment-results}
\end{minipage}

We can observe that using a random scheme, we get the most true positives (correctly targeted users), but at a cost of a very large amount of false positives. However, the linear model, while targeting much fewer people, gets a decent amount of correctly targeted people, while maintain a very low amount of false positives. The breaking cost point represents the minimum cost of a targeting that we can use in order to make profit. Since both true positives and false positives cost this amount, we need that value to be as high as possible. The formula for computing the profit, given a list of targeted people is $profit = \sum_{i \in TP}{revenue(i)} - (TP + FP) * cost $. The revenue vector represents the amount of money that particular user has spent on the last session. The breaking cost point can be obtained by setting the profit to 0, resulting in $BP = \frac{\sum_{i \in TP}{revenue(i)}}{TP + FP}$. In Figure \ref{fig:revenue-per-targeting-cost}, we can observe the linear and log-scale plots for this function, for all 4 methods.

\newpage
\noindent%
\begin{minipage}{\linewidth}
\makebox[\linewidth]{
    \includegraphics[scale=0.5,keepaspectratio]{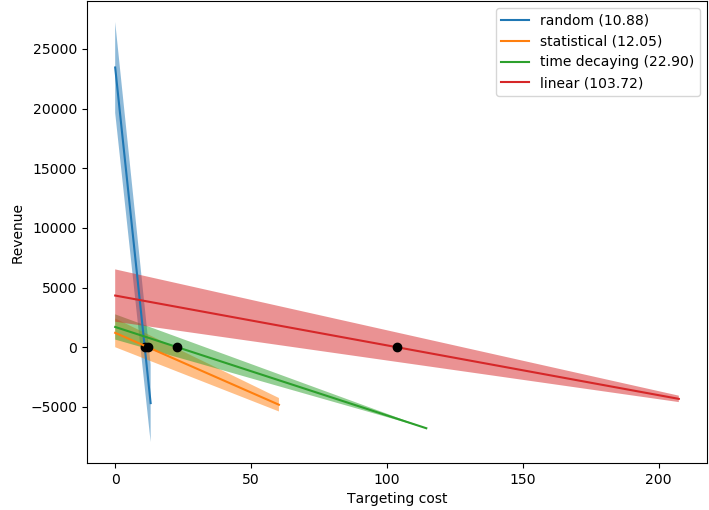}
    \includegraphics[scale=0.5,keepaspectratio]{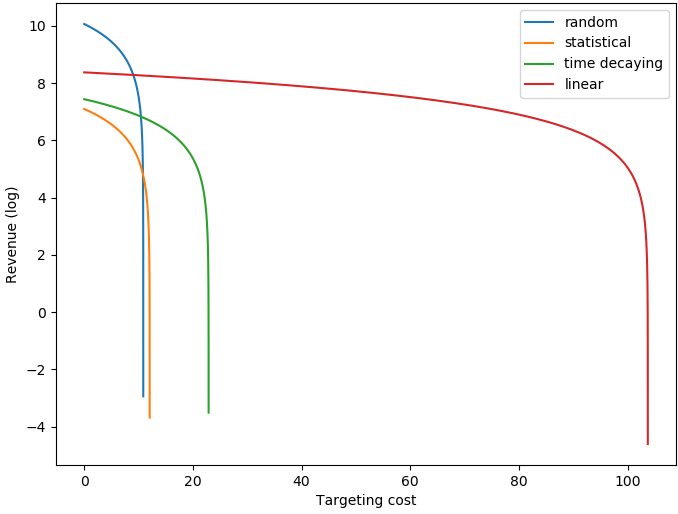}
}
\captionof{figure}{Revenue per targeting cost (linear and log scale)}\label{fig:revenue-per-targeting-cost}
\end{minipage}

We observe that, if the cost of a targeting is very low (close to 0), then the random method would give us the highest profit. However, this cost is seldom so low, because it usually represents the cost of an ad campaign, which has usually much higher values. Also, if we target users that don't want to buy something now, we risk the chance of losing them forever. This is why it is important to only target relevant users, that might convert with a high probability based on its behaviour. As the cost increases, the profit drops drastically for the random scheme. The other 3 methods prove more resilient, with the linear model having the best result with a breaking cost point of over 100.

\section{Conclusions}

In this paper we explain how to use Google Analytics to export relevant data from an E-commerce website and use that data to train a recurrent neural network model in order to predict the probability of an user doing an action. These actions were defined purely using statistics, by taking the most 6 visited paths. However, the most important one is the transaction class. Using the trained models, we provide an targeting experiment, where we'd like to see if the model performs better than a random or a statistical based targeting rule in order to compute the breaking cost point. We observe that the linear attribution model performs the best, much better to all the other 3 methods.

\newpage
\bibliographystyle{unsrt}  

\end{document}